\documentclass[a4paper]{article}
\usepackage{enumitem}
\setitemize{noitemsep,topsep=0pt,parsep=0pt,partopsep=0pt}
\usepackage[a4paper, total={150mm, 220mm}]{geometry}
\usepackage[utf8]{inputenc}
\usepackage{authblk}
\usepackage{natbib}
\usepackage{textcomp}
\usepackage{amssymb}
\usepackage{graphicx}
\usepackage{makecell}
\usepackage{bm}
\renewcommand{\vec}[1]{\mathbf{#1}}
\graphicspath{ {figures/} }
\usepackage{amsmath}
\usepackage{xspace}
\usepackage[T1]{fontenc}

\newcommand{\deepwalk}{\textsc{DeepWalk}}
\newcommand{\wordvec}{\textit{word2vec}}
\newcommand{\nodevec}{\textit{node2vec}}
\newcommand{\strucvec}{\textit{struc2vec}}
\newcommand{\metapath}{\textit{metapath2vec}}
\newcommand{\metapathplus}{\textit{metapath2vec+}}

\newcommand{\skipgram}{Skip-gram}
\DeclareMathOperator*{\argmax}{arg\,\max}
\title{
Tutorial on NLP-Inspired Network Embedding}

\author{
Boaz Shmueli$^1,^2$\\
\vspace*{0.2cm}
$^1$Social Networks and Human-Centered Computing, Taiwan International Graduate Program (TIGP)\\
Institute of Information Science, Academia Sinica\\
\vspace*{0.2cm}
$^2$National Tsing Hua University\\
}
\date{January 2018}

\begin{document}
\maketitle
\setcounter{tocdepth}{1}
% \tableofcontents
% \newpage

\thispagestyle{empty}
\mbox{}
% \newpage
\section{Introduction}
In this  tutorial I cover a few recent papers in the field of network embedding. Network embedding is a collective term for techniques for mapping graph nodes to vectors of real numbers in a multidimensional space. To be useful, a good embedding should preserve the structure of the graph. The vectors can then be used as input to various network and graph analysis tasks, such as link prediction. The papers I discuss develop methods for the online learning of such embeddings. These developments in online learning of network embeddings have major applications for the analysis of graphs and networks, including online social networks. 

Recently, researchers have adapted ideas and techniques from word embeddings to the domain of network embeddings. This tutorial will cover these recent developments. Specifically, I cover the following papers:

\begin{table}[h]
\centering
\label{papers}
\begin{tabular}{p{2.5in}p{1.5in}}
\textbf{Title} & \textbf{Where, Who, When} \\
\deepwalk{}: Online Learning of Social Representations  & KDD \citet{perozzi2014deepwalk}  \\
\hline 
LINE: Large-scale Information Network Embedding &    WWW, \citet{tang2015line}  \\
\hline 
node2vec: Scalable Feature Learning for Networks   &    KDD, \citet{grover2016node2vec}  \\
\hline 
struc2vec: Learning Node Representations from Structural Identity   &  KDD, \citet{ribeiro2017struc2vec} \\
\hline 
metapath2vec: Scalable representation learning for heterogeneous networks 
& KDD, \citet{dong2017metapath2vec} \\
\hline          
\end{tabular}
\caption{Reviewed Papers}
\end{table}

The papers use various methods to sample the nodes and create node contexts. Subsequently, machine learning techniques perform the embedding. The first paper, ``\deepwalk{}: Online Learning of Social Representations'', is a seminal paper that generalizes a well-known NLP word embedding technique, \wordvec{}, to graph theory. The second and third papers, ``LINE: Large-scale Information Network Embedding''  and ``node2vec: Scalable Feature Learning for Networks'', improve upon \deepwalk{} in a substantial way. 

The papers were presented at SIGKDD and WWW conferences (2014, 2015, and 2016) and are highly-cited: 471, 362, and 268 in Google Scholar; 374, 330, and 139 in Microsoft Academic. The latter service is reputed to be more reliable. 

The last two papers reviewed are ``struc2vec:  Learning Node Representations from Structural Identity'' and ``metapath2vec:  Scalable representation learning for heterogeneous networks''.

\begin{figure}
    \centering
    \includegraphics[width=4.5in]{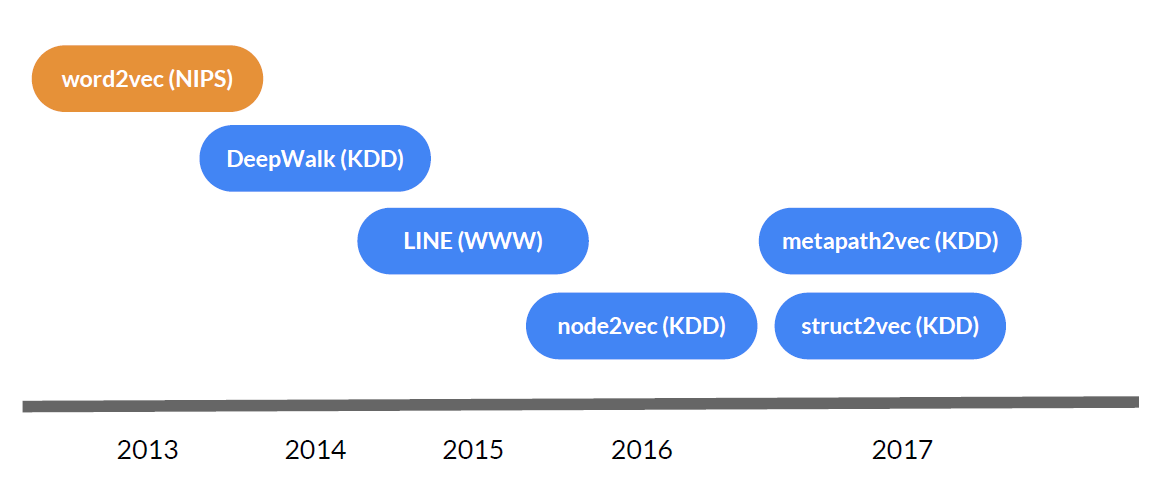}
    \caption[caption]{Paper outline}
    \label{fig:outline}
\end{figure}

Fig. \ref{fig:outline} shows a time outline of the papers reviewed in this tutorial.

The papers use \textit{\wordvec{}}, an algorithm designed for performing word embedding. Thus, I first give a brief introduction to this technique. The original \wordvec{} papers are \citet{mikolov2013efficient} and \citet{mikolov2013distributed}. Since many of the descriptions in these papers are ``somewhat cryptic and hard to follow'' \citep{goldberg2014word2vec}, a small cottage industry of pre-prints explaining \wordvec{} has sprung up, including \cite{rong2014word2vec} and \cite{goldberg2014word2vec}. 

%\subsection{A note on terminology}
%As a new and evolving field, it is expected that there is some confusion regarding terminology. In %the field of NLP, the term \textit{word embedding} is used to describe the mapping of words to real %vectors. Thus, the most appropriate analogy would be  \textit{Graph embedding} is used in %topological graph theory to describe the projection of graphs on surfaces. This has led to the use %of graph embedding in machine learning as

\section{\wordvec{}}
Traditional natural language processing systems treat words as discrete symbols. Words are then represented as one-hot vectors in a very high dimension space. With a vocabulary of $V$ different words, the space is of dimension $V$, and each word is represented by a vector of length $V$ with a single component of 1 and $V-1$ components of 0. 

In contrast, vector space models embed words in a low, $d$-dimension real-numbers vector space such that words which are semantically similar are represented by closer vectors. This embedding is based on the principle of co-occurrence: the assumption that words with related semantic meanings appear close to each other in texts. This assumption is famously summarized by British linguist J. R. Firth's famous statement that ``you shall know a word by the company it keeps'' \citep{firth1957papers}.

\wordvec{} (\citet{mikolov2013efficient}, \citet{mikolov2013distributed}) is a machine learning model for the efficient learning of word embeddings given raw text. There are two versions of \wordvec{}, Continuous Bag-of-Words (CBOW) and \skipgram. Here I focus on \skipgram, since this is the model used by \deepwalk{} and \nodevec{}.

Given a word $w$ we look at its context-words $c \in C(w)$ using a sliding window of size $2 \times k$: the  $k$ words preceding $w$, and the $k$ words following $w$. A typical parameter for this window parameter is $k=5$. When we consider the conditional probabilities $p(c|w)$, the goal is to find the parameters $\theta$ of $p(c|w, \theta)$ such that the following is maximized:
\begin{equation}
    \argmax_\theta 
     \prod_{c \in C(w)} p(c|w;\theta)
\end{equation}
For the entire corpus $T$ we have:
\begin{equation}
    \argmax_\theta \prod_{w \in T} 
    \left( \prod_{c \in C(w)} p(c|w;\theta) \right)
\end{equation}
which we can also write as 
\begin{equation}
    \argmax_\theta \prod_{(w, c)\in D} p(c|w;\theta) 
    \label{eq:node2vec}
\end{equation}
where $D$ is the set of all the $(w, c)$ pairs of (word, context-word). In a corpus $T$ with a vocabulary of $V$ words, we have approximately $2 \times k \times V$ such pairs. 

\begin{figure}
    \centering
    \includegraphics[width=3.5in]{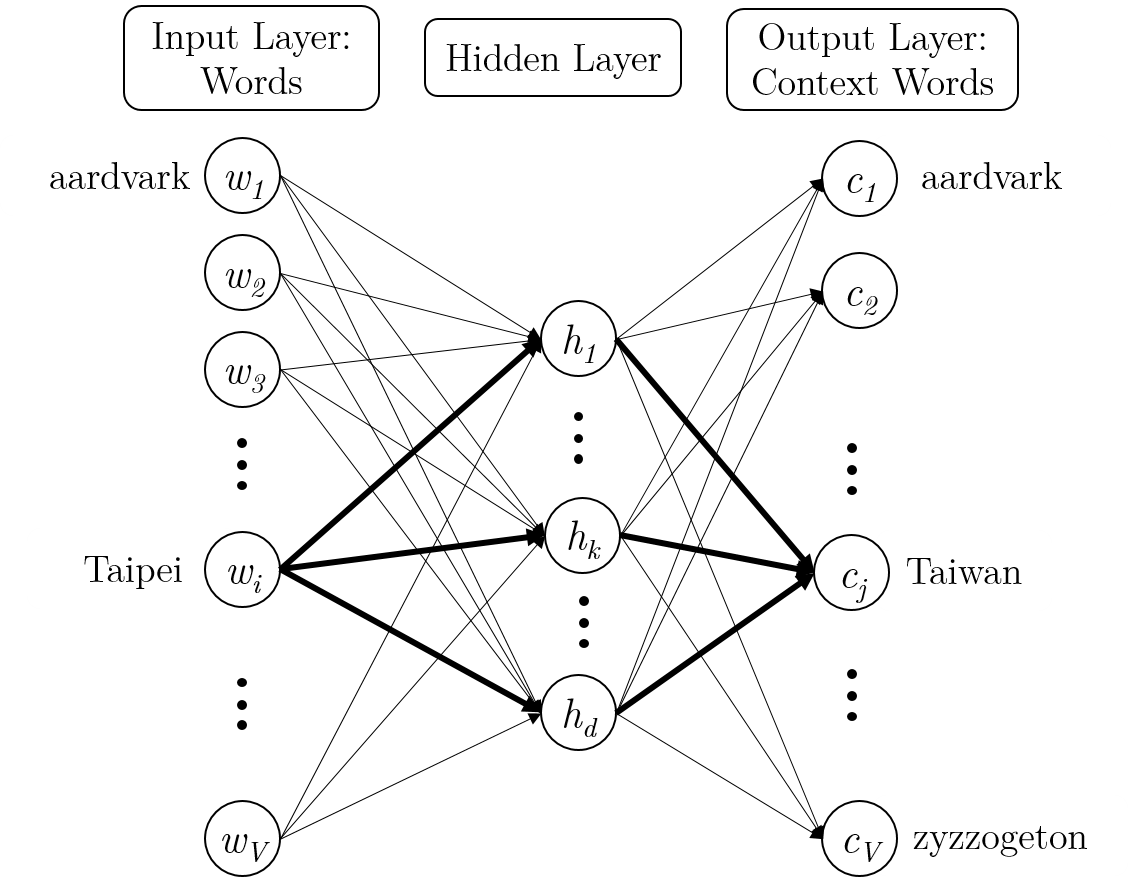}
    \caption[caption]{\wordvec{} neural network}
    \label{fig:word2vec}
\end{figure}

The approach taken by  $word2vec$ is to parameterize the model using a classical soft-max neural network:
\begin{equation}
    p(c|w;\theta) = \frac
    {e^{\vec{v_c} \cdot \vec{v_w}}}
    {\sum\limits_{c^\prime \in C} e^{\vec{v_c^\prime} \cdot \vec{v_w}}}
\end{equation}
where $\vec{v_w}$ and $\vec{v_c}$ are the vector representations of the words and the context-words, both in $\mathbb{R}^d$. 
The input layer thus consists of $V$ neurons where each word is represented by a one-hot vector. The output layer has $V$ neurons, one for each context word, and acts as a soft-max classifier. The hidden layer has $d$ neurons, where $d$ is the size of the embedding space. This is shown in Fig. \ref{fig:word2vec}.

As is usually the case, the objective function is easier to optimize after taking its log. This step is omitted here for sake of brevity. Following the optimization, the $V \times d$ weights of the $d$ neurons of the hidden layer serve as the vector representations. The hidden layer basically serves as an auto-encoder that produces the word embeddings. The outputs from the final layer are not needed.

Due to the time complexity of the training (there are $V \times V \times d$ parameters to be computed), it is impractical to scale the training of the network to very large corpora. The authors describe two tricks for better computation efficiency. The first is hierarchical softmax, an efficient way of computing softmax (\citet{mnih2008probabilistic} and \citet{morin2005hierarchical}). Hierarchical softmax reduces the complexity of computation per training instance per context word from $O(V)$ to $O(\log V)$. The second is negative-sampling, where only a sample of the output vectors are updated per iteration. These two improvements make it possible to train the model on very large amounts of text.
\footnotetext{The figures in this tutorial are drawn by the author, unless otherwise stated}

\subsection{A Note on Terminology}
There is some confusion regarding terminology in this cutting-edge field, especially between \textit{graph embedding} and \textit{node embedding}. It would perhaps be more consistent to use \textit{node embedding} for the mapping of nodes to vectors, in the same way that \textit{word embedding} is used in NLP. Indeed, some authors use this term. Others, however, use graph embedding. This is confusing since graph embedding is also used to denote the mapping of entire graphs into a single vector (in much the same way that \textit{document embedding} maps an entire document into a vector and thus allows to find distances between documents). Another factor adding to the confusion is that graph embedding is a well establish topic in the mathematical field of topological graph theory. 

In this tutorial I chose to use yet another term, \textit{network embedding}, to describe the mapping of the nodes to vectors, as this currently seems to be the most popular term in the machine learning community.

\section{\deepwalk{}}

\wordvec{} is a huge success, and inspires many derivative works in related fields. One of the most intriguing developments is \deepwalk{} \citep{perozzi2014deepwalk}, a new online learning method for embedding of graphs, with an emphasis on graphs representing social networks.

The problem of network embedding can be formalized as follows: given a graph $G(V,E)$ with a set $V$ of nodes and a set $E$ of edges, compute a mapping $f(v \in V) \longrightarrow \mathbb{R}^d$ such that the structure of $G$ is preserved as much as possible. 
Vast literature in the field covers multiple techniques for attacking the problem of network embeddings \citep{fouss2016algorithms}. For example, \emph{spectral methods} use the eigenvectors of various graph matrices to compute the embeddings, thus offering exact solutions following a closed-form formulation. Another approach is to use the analogy between nodes and edges to physical spring networks, resulting in \emph{force-directed methods}. 

\deepwalk{}, on the other hand, gets its inspiration from NLP and generalizes the \wordvec{} model to graphs using an analogy between documents and graphs, as shown in Table \ref{table:analogy}: a document is seen as a graph of words. In the \wordvec{} model, words are used to estimate the likelihood of the nearby context-words in the sentence. 
In much the same way, \deepwalk{} uses nodes to estimate the likelihood of nearby nodes in the graph. To paraphrase J. R. Firth, ``you shall know a node by the company it keeps''. The nearby nodes are sampled using random walks. The assumption is that sampling from multiple random walks captures the structure of the graph. 
\begin{table}[h]
\centering
\begin{tabular}{ll}
NLP          & Graph Theory \\ \hline
Corpora/Documents      & Networks/Graphs\\
Words        & Nodes/Vertices\\
Word co-occurrence & Node neighborhood\\
Sentences & Random walks
\end{tabular}
\caption{Analogy between NLP and Graph Theory}
\label{table:analogy}
\end{table}

The \deepwalk{} algorithm generates random walks for each node $v \in V$. Each random walk starts from the origin node $v$ and then advances to a node uniformly selected from its immediate neighbors. The length of the walks is $T$. Thus, each such random walk generates an ordered list of nodes:
\begin{equation}
    RW(v) = (v=u_1, u_2, ..., u_{T}).
\end{equation}
Each walk is a ``sentence'' of nodes. Similarly for the words in \wordvec{}, for each node $u_{i}$ within this ``sentence'', the algorithm then looks at a window of $k$ neighboring nodes before and after $u_j$, which comprise the set of our ``context-nodes'':
\begin{equation}
    C(u_{i}) = \{u_{i-k}, ..., u_{i+k}\}
\end{equation}
We would then like to find $\theta$ that maximizes:
\begin{equation}
    \argmax_\theta 
     \prod_{u \in RW(v)} p(u^\prime \in C(u)|u_;\theta)
\end{equation}
For the entire graph $G$ we have:
\begin{equation}
    \argmax_\theta \prod_{v \in G} 
    \left( \prod_{u \in RW(v)} p(u^\prime \in C(u)|u_;\theta) \right)
\end{equation}
which we can also write as 
\begin{equation}
    \argmax_\theta \prod_{(u^\prime, u)\in D} p(u^\prime|u;\theta) 
    \label{eq:deepwalk}
\end{equation}
where $D$ is the set of all (node, context-node) pairs discovered in all the random walks of all the nodes. 
To improve sampling, $\gamma$ random walks per node $v$ are performed. $\gamma$, $T$, and $k$ are all hyper-parameters of the algorithm. In total, $\gamma  \times |V|$ random walks  of length $T$ are generated, and there are approximately $\gamma \times |V| \times T \times k$ (node, context-node) pairs.

As can be seen, \deepwalk's Equation \eqref{eq:deepwalk} is equivalent to \nodevec{} \skipgram{}'s Equation \eqref{eq:node2vec}, and indeed the same neural network is used to compute the embeddings for the nodes, with some mandatory adjustments (for example, the input layer is the one-hot representations of the nodes). \subsection{Experiments}
The original motivation for \deepwalk{} is solving the following social graph multi-label classification problem: we are given a graph $G(V,E)$ of a social network and node attributes 
$X \in \mathbb{R}^{|V| \times S}$ (where $S$ is the size of the feature space). There is also a set of labels $\mathcal{Y}$. Some of the nodes are labeled with $y \in \mathcal{Y}$. The task is to predict the labels of the other (unlabeled) nodes.\footnote{In fact, the paper mentions that the label output is $Y \in \mathbb{R}^{|V| \times \mathcal{Y}}$. I found this confusing, if not wrong, since the labels are not real numbers but belong to a discrete set of labels.}

This problem is known as a \textit{relational classification problem}. Traditional approaches use the graph
structure for classification. In this paper, we first learn the embedding of the nodes 
$X_E \in \mathbb{R}^{|V| \times d}$, 
and these are then treated as additional features. $X_E$ is thus combined with the attributes $X$ as input to any standard classification algorithm.

The authors compared the performance of \deepwalk{} with five other baseline methods:
\begin{itemize}
\item Spectral Clustering of the normalized graph Laplacian \citep{tang2011leveraging} 
\item Modularity Matrix \citep{tang2009relational}
\item $k$-means clustering  of the adjacency matrix  \citep{tang2009scalable}, 
\item Weighted-vote relational neighbor \citep{macskassy2003simple}, 
\item Na\"ive majority
\end{itemize}

The classification method used with \deepwalk{} is a one-vs-rest logistic regression, and the parameters selected are $d=128$ (number of dimensions), $T=50$ (length of random walks), $w=10$ (window size), $\gamma = 80$ (number of random walks per node).

Three test datasets are used for the multi-label classification task: \textsc{BlogCatalog}, \textsc{Flickr}, and \textsc{YouTube}. The results show that \deepwalk{} has strong performance that is almost as good as or exceeds the leading method, Spectral Clustering. In addition, it needs relatively few labels to perform well. It also has the advantage that it is computationally feasible to perform this embedding on huge networks such as \textsc{YouTube}, which is computationally unfeasible for spectral clustering.

%%%
%%% LINE
%%%

\section{LINE}
While the analogy of graphs and texts is useful, networks do not possess the linear property of text. So while the neighborhood of a word can be quite accurately sampled using a sliding window, social and other large networks are not linear, and thus sampling their structure is more challenging.
Following the introduction of \deepwalk{}, many other researchers have started to use similar NLP-inspired methods for network embedding. One of the problems with \deepwalk{} is that it uses unbiased random walks for generating the node contexts. In that way, it is similar to a depth-first search (DFS). 
The work by \cite{tang2015line} tries to solve this issue by preserving first-order and second-order node proximities.

\begin{figure}[h]
    \centering
    \includegraphics[width=2in]{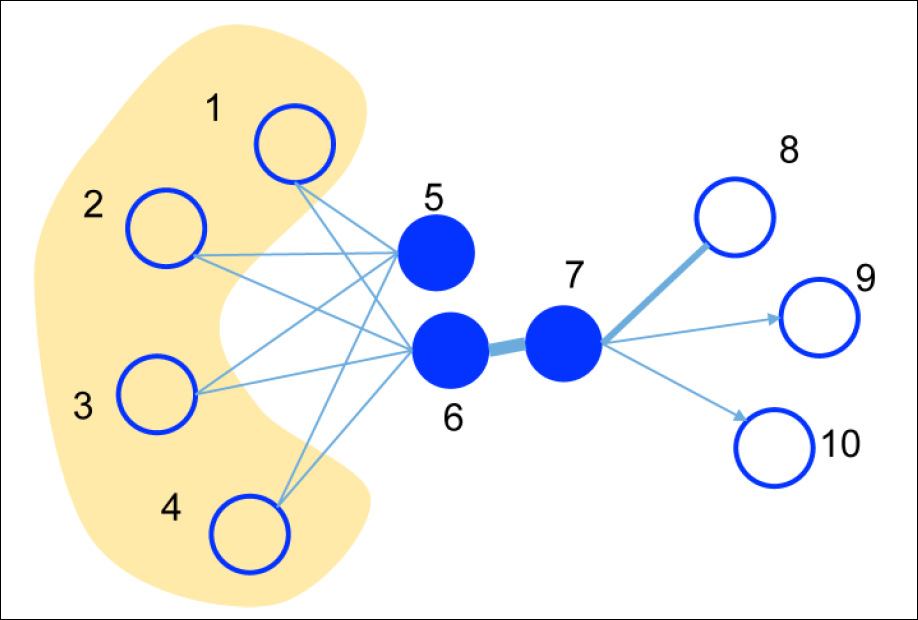}
    \caption{First-order and second-order proximities}
    \label{fig:line}
\end{figure}

The first-order proximity of two nodes is defined to be the weight of the edge between them (1 for unweighted graphs), and zero if they do not share an edge. For example,  nodes 6 and 7 in Fig. \ref{fig:line} should be embedded closely since they are connected by a ``heavy'' edge (compared to nodes 7 and 8, for example). 

The second-order proximity is defined by the common neighborhood of two nodes. Referring back to Fig. \ref{fig:line}, nodes 5 and 6 have a high second-order proximity since they share a lot of neighbors, even thought they are not directly connected. Thus, they should also be embedded closely together in the embedding space.

In mathematical terms, the first order proximity between any two nodes $v_i$ and $v_j$ is defined by LINE to be the following joint probability:
\begin{equation*}
    p_1(v_i, v_j)=\frac{1}{1+exp(-\vec{v_i}^T \cdot \vec{v_j})}
\end{equation*}
where $\vec{v}$ is the vector representation of node $v$ in $\mathbb{R}^d$.
Ideally, this probability would be equal to the empirical probability induced by the edge weights:
\begin{equation*}
    \hat{p_1}(v_i, v_j)=\frac{w_{ij}}{W}
\end{equation*}
where $w_ij$ is the weight of the edge between the two nodes and $W$ is a normalization factor.
The embedding should try and minimize the distance between the two distributions $p_1(\cdot, \cdot)$ and $\hat{p_1}(\cdot, \cdot)$. LINE uses KL-divergence to measure distance between distributions, and so we get that the distance to optimize is:
\begin{equation*}
 O_1 = - \sum_{(i,j) \in E} w_{ij} \log (p_1(v_i, v_j))
\end{equation*}
For the second-order proximity, the authors choose the neighbors of each node to provide ``context'' in the \wordvec{} sense. The development for this is quite tedious, so we will skip it here. The reader is referred to Section 4.1.2 in \cite{tang2015line}. The result is the following objective function for the second order proximity:
\begin{equation*}
 O_2 = - \sum_{(i,j) \in E} w_{ij} \log (p_2(v_j| v_i))
\end{equation*}
Where, similarly to the \skipgram{} model, the conditional probability between two nodes is given by:
\begin{equation*}
 p_2(v_j|v_i)=\frac{\exp ( \vec{v_j}^\prime \cdot \vec{v_i})}{ \sum_{k=1}^{|V|}   \exp ( \vec{{v\prime}_k^T} \cdot \vec{v_i})}
\end{equation*}
The second-order proximity thus defines a conditional probability $p(v_j|v_i)$ over all other context nodes $v_k$.

\subsection{Embedding and Experiments}
In LINE, the embedding for the first-order and second-order proximities (i.e. maximizing for $O_1$ and $O_2$) is done separately. Solving for $O_2$ is done using asynchronous stochastic gradient algorithm (ASGA) with various performance optimization tricks, including edge sampling.
The vectors from the two models are then concatenated for each node. As the authors note, this approach is not very ``principled'', and a better approach that would minimize $O_1$ and $O_2$ simultaneously is more appropriate.

LINE is compared against baseline methods for performing various tasks. The data includes a Language Network from the entire English Wikipedia, Social Networks (Flickr and YouTube), and Citation Networks. Algorithms for comparison included graph embedding using matrix factorization, \deepwalk{}, and various variations of the LINE algorithm. The results show clearly that the LINE network embedding that include both first-order and second-order proximities outperform all other methods in classification tasks.

\section{\nodevec{}}

Following \deepwalk{} and LINE, \citet{grover2016node2vec} had the insight that a better capture of a node's neighborhood can be achieved by carefully biasing the random walks. This leads to latent representations that better capture the network structure. \nodevec{} indeed achieves excellent performance in multiple social graph tasks.

Assuming that we want a sample of $k$ nodes from the neighborhood of each node. Two extreme strategies are Breadth-First Sampling (BFS) and Depth-First Sampling (DFS). BFS samples the immediate $k$ neighbors of the node, and thus helps in understanding its local structure. DFS samples $k$ increasingly distant nodes, and thus identifies a more global community structure. A rich sample that combines the properties of BFS and DFS is the intuition behind the creation of the biased random walk.

Let us recall that the random walk proposed by \deepwalk{} had the following uniform probability of advancing from node $u_i$ to node $u_{i+1}$:
\begin{equation}
    P(u_{i+1} = v | u_i = w) = 
  \begin{cases} 
     \frac{1}{d(w)} & \text{if $(v,w) \in E$}\\
     0 & \text{otherwise}
   \end{cases}
\end{equation} 
where $d(w)$ is the degree of node $w$.

In contrast, the random walk proposed by \nodevec{} has two parameters that control the walk, $p$ and $q$. Assume that in the random walk we just advanced from node $t$ to node $v$. There are three possibilities for the next node in the walk: 
we can (1) go from $v$ back to $t$, (2) advance to a third node $x$, a common neighbor of both $t$ and $v$,
or (3) advance to a third node $x$, a neighbor of $v$ but not of $t$. 
The parameters $p$ and $q$ control the probabilities for each of these types of transitions.

\begin{figure}[h]
    \centering
    \includegraphics[width=2in]{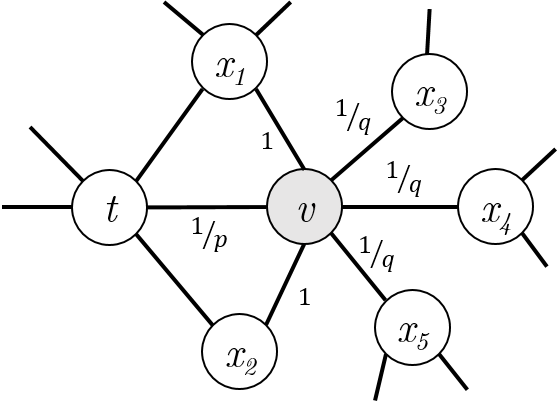}
    \caption{Unnormalized transition probabilities from node $v$, after just transitioning from node $t$}
    \label{fig:transprops}
\end{figure}

More specifically, assuming identical edge weights, the \textit{unnormalized} transition probability from node $v$ to node $x$, having arrived from node $t$, is:
\begin{equation}
    \pi_{vx}^t=
  \begin{cases} 
     \frac{1}{p} & \text{if $d(t,x) = 0$}\\
     1 & \text{if $d(t,x) = 1$}\\
     \frac{1}{q} & \text{if $d(t,x) = 2$}
   \end{cases}
\end{equation} 
where $p$ is the \textit{return parameter}, which controls the likelihood of the walk to backtrack to the previous node; and $q$ is the \textit{in-out parameter}, which decides whether to favor nodes closer to $t$. See Fig. \ref{fig:transprops} for an example of the transition properties.
By controlling these two parameters, the randoms walks can achieve a balance of the benefits of both BFS and DFS, and thus more accurately represent the local and global graph properties. 
The rest of the exposition in \citet{grover2016node2vec} follows similar path to \deepwalk{}, i.e., it uses the \wordvec{} neural network model with a hidden layer that calculates the latent vector representations. Optimization is done using stochastic gradient descent (SGD). The algorithm also borrows \wordvec{}'s negative sampling trick to achieve scaling to graphs with millions of nodes.

\subsection{Edge Features}
In addition to biased random walks, one of the interesting contributions of \nodevec{} is the 
extension of feature learning to edges. For this goal, a binary operator $\circ$ is defined on the vector representations $f(v)$ and $f(u)$ of \textit{any} two nodes $v$ and $u$, with the aim of generating a vector representation of the pair $(u,v)$ such that $g(u,v) : V \times V \longrightarrow \mathbb{R}^{d^\prime}$. Several choices for this binary operators where $d=d^\prime$ are shown in Table \ref{table:edge}. The operator is defined whether or not the edge actually exists in the graph, paving the way for the use of this representation in a link prediction task. 

\begin{table}[h]
\centering
\begin{tabular}{ll}
Operator          & Definition  \\ \hline
Average     & $\frac{f_i(u) + f_i(v)}{2}$ \\
Hadamard    & $f_i(u) * f_i(v)$ \\
Weighted-L1    & $|f_i(u) - f_i(v)|$ \\
Weighted-L2    & $|f_i(u) - f_i(v)|^2$ \\
\hline
\end{tabular}
\caption{Different operators showing the $i$th component of $g(u,v)$}
\label{table:edge}
\end{table}

\subsection{Experiments}
The authors test \nodevec{} with two tasks. The first is a multi-label classification task, similar to the one used with \deepwalk{}. In this task, the node feature representations are input to a one-vs-rest logistic regression classifier. Three datasets are tested: \textsc{BlogCatalog} (10,312 nodes, 39 different labels); \textsc{Protein-Protein Interaction (PPI)} (3,890 nodes, 50 different labels); and \textsc{Wikipedia} (4,777 nodes, 40 different labels). 

The baseline algorithms tested against are:
\begin{itemize}
\item \deepwalk{}
\item Spectral Clustering of the normalized graph Laplacian \citep{tang2011leveraging} 
\item LINE
\end{itemize}
With the right selection of $p$ and $q$, \nodevec{} outperforms all other contenders. The performance gain for BlogCatalog and Wikipedia is a staggering 22\%.

In addition to the classification task, the authors also experiment with a link prediction task using the various edge features operators that are described above. Three datasets are tested: \textsc{Facebook} (4,039 nodes; 88,234 edges); \textsc{PPI} (19,706 nodes; 390,633 edges); \textsc{arXiv} (18,722 nodes; 198,110 edges). They compare \nodevec{} performance against the baseline algorithms using the different operators in Table \ref{table:edge}, as well as against the standard link prediction scores: Common Neighbors, Jaccard's coefficient, Adamic-Adar Score, and Preferential Attachment. The \nodevec{} method with the Hadarmard operator outperforms all the other methods, in some cases with impressive improvements.

\section{\strucvec{}}
Another interesting network embedding work is \strucvec{} \citep{ribeiro2017struc2vec}, which focuses on the role of nodes in a network. Nodes in networks have specific roles. These roles can be identified through structural identity. For example, in an airport network, some nodes serve as hubs. Two hubs can be far away (hop-wise), but still have structural similarity. In social networks, roles of users are can also be identified by the structure. For example, in a graph representing a company structure, mid-level managers have a typical structure within the graph.
\begin{figure}
    \centering
    \includegraphics[width=2in]{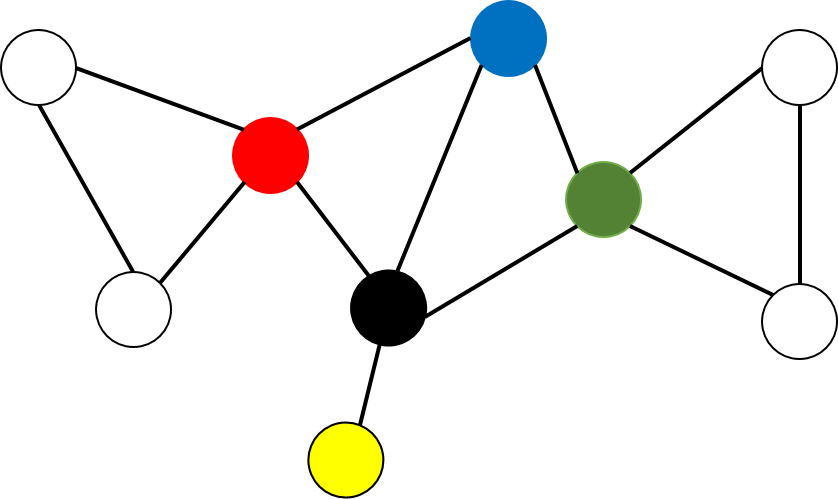}
    \caption{Node roles in graphs}
    \label{fig:automorphism}
\end{figure}
Consider the graph in Fig. \ref{fig:automorphism}. The red and green nodes are structurally equivalent. They belong to an isomorphism. The blue and black nodes, while not structurally \textit{equivalent}, are structurally \textit{similar}. 

\strucvec{} is a framework for representations based on structural similarity.  The goal of \strucvec{} is to preserve the identity of the nodes' structure when projecting them into Euclidean space, even if they are not close. Embeddings such as \deepwalk{} and \nodevec{} capture neighborhood relations. 
However two nodes that are structurally similar but very distant will not be close in the vector space. In addition, nodes that are close to each other in the graph can be structurally dissimilar, and thus should not be close in the Euclidean space. 
This is the problem that \strucvec{} is trying to solve. In other words, the embedding done by the former methods depends on the hop distance between nodes. \strucvec{} does not take into account this distance. 

To compute the representations, \strucvec{} builds a special graph - the \textit{context graph} - that represents structural similarities between nodes. The goal is to create a context graph where nodes are close to each other if they are structurally similar. Once the context graph is constructed, the embedding is done once again using the \wordvec{} algorithm. That is, random walks are performed in the context graph in order to build "sentences", followed by online learning using \wordvec{}'s \skipgram{} algorithm. Thus, the main contribution of \strucvec{} is the construction of the context graph.

Given the original graph $G(V, E)$ with diameter $K$, the context graph $M$ is a multi-layer graph with $K+1$ layers. Each layer includes all the nodes in $G$. Within each layer, weighted edges represents structural similarity between nodes. Edges also exist between the corresponding nodes of each layer. We will now describe the four steps used in \strucvec{}. Our emphasis is on the construction of the context graph, since this is the main contribution of the paper.

\subsection{Step 1: Computing Structural Similarity}
The first step looks at the structural similarity between nodes. For each node $v$, we look at the $N_k(v)$, which is the set of nodes which are $k$-distant from u ($N_0(v) = {v}$). These form ``rings'' around each node. For each such ring, we look at the ordered degree sequence $DS(N_k(v))$ of the nodes participating in the ring. 

Referring again to Fig. \ref{fig:automorphism}, let's focus for example on the $black$ node. When $k=0$, $N_0(black) = \{black\}$, and its degree is 4, thus $DS(N_0(black)) = (4)$. Moving to $k=1$, we look at the nodes whose distance to $black$ is $1$. We see that:
\begin{equation*}
    N_1(black) = \{red, blue, green, yellow\}
\end{equation*}
and their degrees are 4, 3, 4, and 1, respectively. Thus, the ordered degree list is:
\begin{equation*}
    DS(N_1(black)) = (4, 4, 3, 1)
\end{equation*}

Once we compute $DS(N_k(v)) \quad (k=0, 1, ..., K)$, for all $v$, we can measure the structural similarities between every pair of nodes $u, v$ by comparing their degree sequences. To measure the distance between two degree sequences $A = DS(N_k(u))$ and $B = DS(N_k(v))$, and noting that these sequences are not necessarily of the  same length, the authors suggest using Dynamic Time Warping (DTW), a technique to match elements of two sequences of different lengths. The technique minimizes the sum of the distances between matched elements $a \in A$ and $b \in B$. The distance between two elements is designed in \strucvec{} so that node degrees of 1001 and 1002 are more similar than node degrees of 1 and 2 by using $\big(max(a,b)/min(a,b)\big)-1$ as the distance between the two matched degrees in the sequences.

Now that the distance between two degree sequences is defined, the \textit{structural distance}
 between two nodes is defined hierarchically as:
\begin{equation*}
    f_k(u,v)=f_{k-1}(u,v)+g\big(DS(N_k(u)), DS(N_k(v))\big).
\end{equation*}
$g$ is the distance function between two degree sequences as described above.

\subsection{Step 2: Constructing the Context Graph}
Now that we defined a structural distance between nodes, we can move on to build the context graph. The context graph  $M$ is a multilayer graph with $K$ layers. Each layer is a complete graph consisting of all the nodes $u \in V$ in the original graph $G$. Thus, each node $u \in V$ is represented in $M$ by $K+1$ nodes $u_k \quad (k=0, 1, ..., K$), one in each layer. 

\begin{figure}[h]
    \centering
    \includegraphics[width=3.5in]{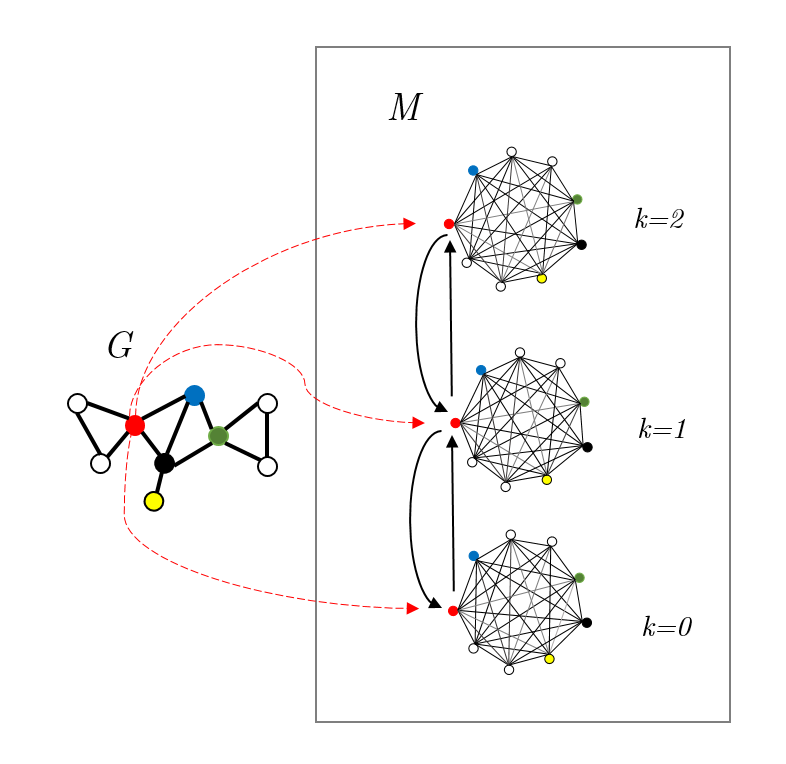}
    \caption{Node roles in graphs}
    \label{fig:multilayer}
\end{figure}

To illustrate, Fig. \ref{fig:multilayer} shows the first three layers of the context graph $M$ for our example graph $G$. Each of the nodes in the original graph $G$ is represented in each of the layers in $M$. The dashed arrows show how the $red$ node is replicated in each layer. Other nodes are similarly replicated. Within the $k$th layer, the weight of an edge between nodes $u_k$ and $v_k$ is a function of the structural distance $f_k(u,v)$. Thus, each layer is a complete sub-graph with weighted, undirected edges that correspond to the structural similarity between the nodes.

Assuming $u_k$ and $v_k$ in layer $k$ are the nodes representing $u$ and $v$, the weight of their connecting edge is calculated as follows:
\begin{equation*}
    w(u_k,v_k)=e^{-f_k(u,v)}, \quad k=0, 1, ..., K
\end{equation*}
Nodes that are structurally similar will have larger weights within the multiple layers of $M$.

Edges also exist between the different layers, but they are directed and exist only between the corresponding nodes $u_0, ..., u_k, u_{k+1}$ that represent the same node $u$ in $G$. Each node is linked to the corresponding node in the layer just above (for $k<K$) and just below (for $k>0$). These directed edges are also weighted. The weights between the same node in different layers is given by:
\begin{equation}
    w(u_k, u_{k+1}) = log(\Gamma_k(u)+e), \quad k=0, ..., K-1
    \label{eq:moveup}
\end{equation}
\begin{equation}
    w(u_k, u_{k-1}) = 1, \quad k=1, ..., K
    \label{eq:movedown}
\end{equation}
where $\Gamma_k(u)$ is the number of edges connected to $u_k$ whose weight is larger than the average edge weight in layer $k$, and is thus a measure of the number of similar nodes in layer $k$. If $u_k$ has many similar nodes, the weight going to the upper level will be larger. In a higher level, the number of similar nodes can only decrease. 

Fig. \ref{fig:multilayer} shows these up/down links between the various representations of the red node in the multi-layer graph. Similar links exist also for the other nodes, but they are not shown in the figure.

\subsection{Step 3: Generate Context}
Once the context graph $M$ is constructed, the rest of the procedure is similar to \deepwalk{}. Random walks are generated in $M$ to create the context of nodes. A hyperparamter $q$ determines whether the walk will change a layer or stay within the layer. The weights of the edges determine the probabilities of advancing to the next nodes.
With probability $q$ the walk stays within the same node, and in that case the probability of moving from $u$ to $v$ is given by
\begin{equation}
    p_k(u,v)=\frac{e^{-f_k(i,v)}}{Z_k(u)}
\end{equation}
$Z_k(u)$ is a normalization factor. 
Thus, in each step, the walk will prefer to walk to nodes which are structurally similar.
With probability $1-q$, the walk will move up or down a layer to its corresponding node in layer $k-1$ or $k+1$ according to the weights given in \eqref{eq:moveup} and \eqref{eq:movedown}.

\subsection{Step 4: Learn Representations}
In \strucvec{} the \skipgram{} approach is  used by generating sets of independent and relatively short random walks in $M$. Multiple random walks are generated for each node, starting with layer $0$. These are the ``sentences''. These sentences are then used as input to the \wordvec{} algorithm to train a neural network and learn the latent representing of the nodes by maximizing the probability of nodes within a context. 

\subsection{Optimization}
$M$ is a huge graph. It has $(K+1)n$ nodes and $\frac{1}{2}(K+1)\big(n(n-1)\big)+2nK$ edges\footnote{The paper has a typo in these numbers, which I verified by corresponding with one of the authors (Savarese). In section (3.2) of the paper the number of nodes in  $M$ is listed as $Kn$ nodes and the number of edges is $\frac{1}{2}K\big(n(n-1)\big)+2n(K-1)$. Thus, the authors mistakenly use $K$ layers instead of $K+1$ layers.}. To reduce the time to generate and store the multi-layer graph and context for nodes, the authors propose three optimizations:
\begin{itemize}
    \item Reduce the length of degree sequences by compressing them into a list of 2-tuples $(d, count)$, which means that the sequence has $count$ nodes of degree $d$ for each such tuple.
    \item Reduce the number of edges in the multilayer graph (only $log(n)$ neighbors per node).
    \item Reduce the number of layers in the multilayer graph.
\end{itemize}
These optimizations enable the \strucvec{} algorithm to scale quasi-linearly, and the authors were able to analyze networks with millions of nodes.
\begin{figure}[h]
    \centering
    \includegraphics[width=3.5in]{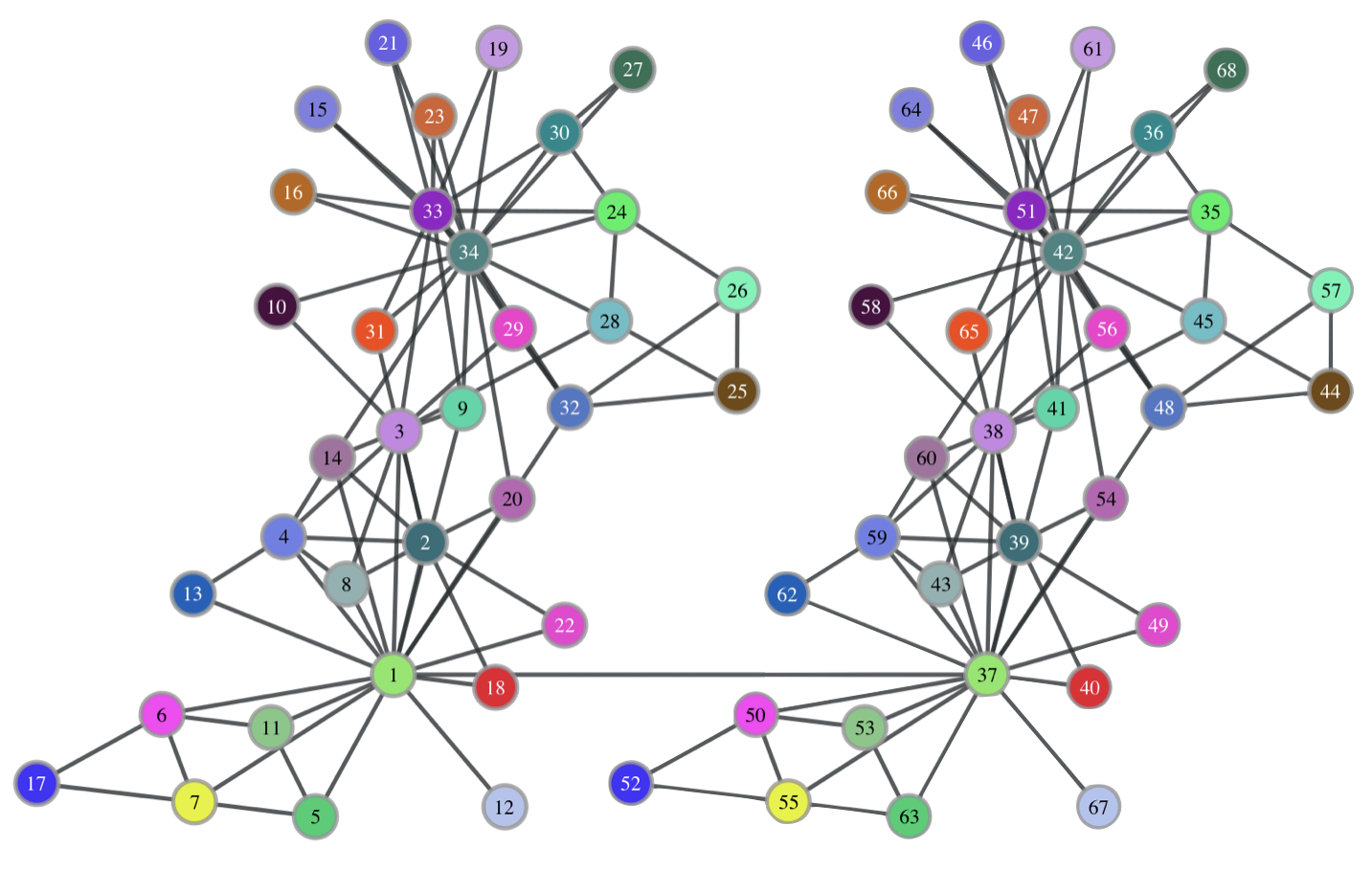}
    \caption{Mirrored Karate network (from \citet{ribeiro2017struc2vec}}
    \label{fig:karate}
\end{figure}

\begin{figure}
    \centering
    \includegraphics[width=3.5in]{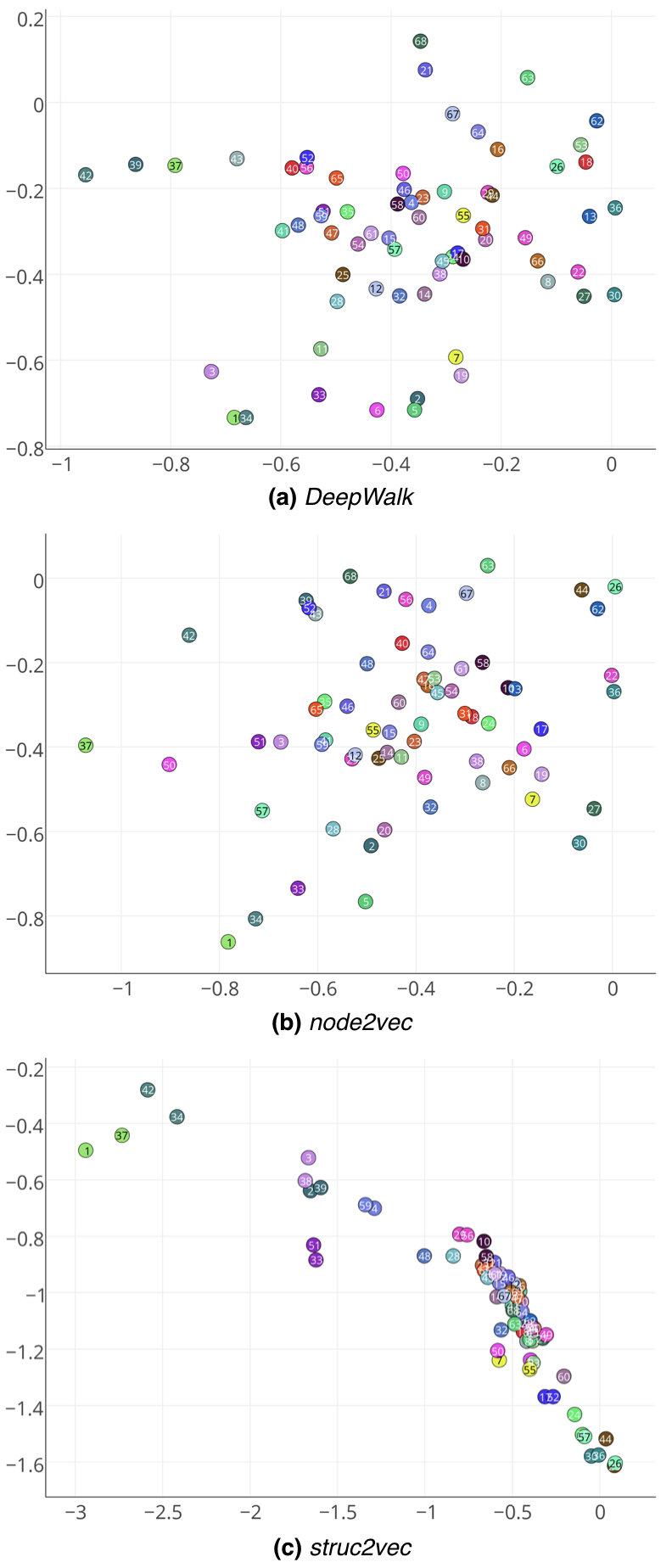}
    \caption{Comparison of Node Representations (from \citet{ribeiro2017struc2vec}}
    \label{fig:struc2compare}
\end{figure}

\subsection{Experiments}
The authors ran different experiments which demonstrated the superiority of \strucvec{} for the task. In one of the experiments, they created a graph composed of two copies of  Zachary's Karate network \citep{zachary1977information}, connected by a single edge, as shown in Fig. \ref{fig:karate}. The result of the \strucvec{} embedding was compared to \deepwalk{} and \nodevec{}. A result of the embedding is shown in Fig. \ref{fig:struc2compare}. Since the Karate network was duplicated, the resulting mirror network has two identical nodes for each role. These was captured by \strucvec{}, as can be seen by the pairs of nodes embedded closely in part (c) of the figure. The top embeddings, for (a) \deepwalk and (b) \nodevec{} do not capture the structural equivalence and mostly focus on the graph distances.

\section{\metapath{}}

Finally, we  look into recent work by \cite{dong2017metapath2vec} which embeds \textit{heterogeneous} networks. 

\begin{figure}[h]
    \centering
    \includegraphics[width=4.5in]{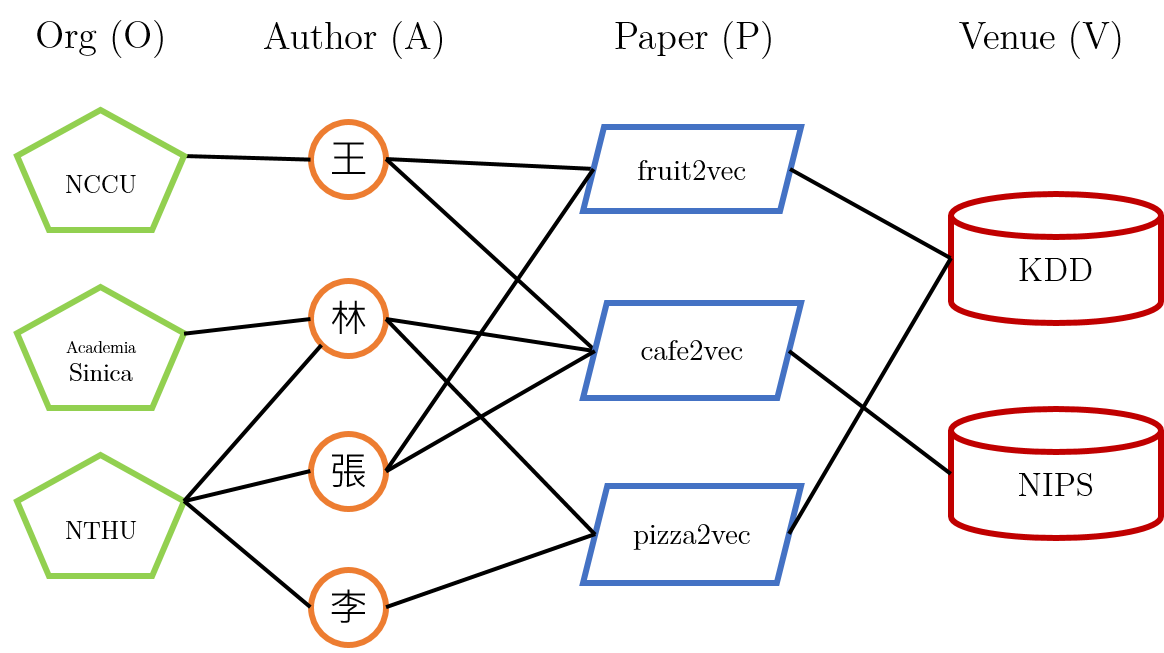}
    \caption{An example heterogeneous academic network, with four types of nodes, from left to right: Organizations (O), Authors (A), Papers (P), and Conference Venues (V)}
    \label{fig:hetero}
\end{figure}

In a heterogeneous graph nodes can represent different entities. A classic example is an academic network, where nodes can represent researchers, organizations, papers and conference venues, and edges represent various relations between the entities. 
For example, a paper is connected to its author(s), and also to the conference venue where it was presented. Fig. \ref{fig:hetero} shows an mock academic network.

The methods we have seen so far (\deepwalk{}, LINE, \wordvec{}) assume a homogeneous network, where there is only one kind of node (e.g., a person in a social network). The contexts generated for the nodes, and the resulting embedding, do not take into account the different types of nodes and relationships between them. To solve this problem, the authors suggest using a method called \metapath{}, where random walks will be biased by using meta-paths. A meta-path is a predefined composite relation between nodes. For example, in the context of an academic network, the specific relation Author-Paper-Author in the network defines the notion of co-authorship. Two authors that are connected in this way are co-authors of the same paper (note that authors can also be connected via an organization, but that implies a different relationship). The idea of the meta-path is not new, but here the authors are using it to create random walks, thus generating carefully biased contexts for nodes.

\begin{figure}
    \centering
    \includegraphics[width=4.5in]{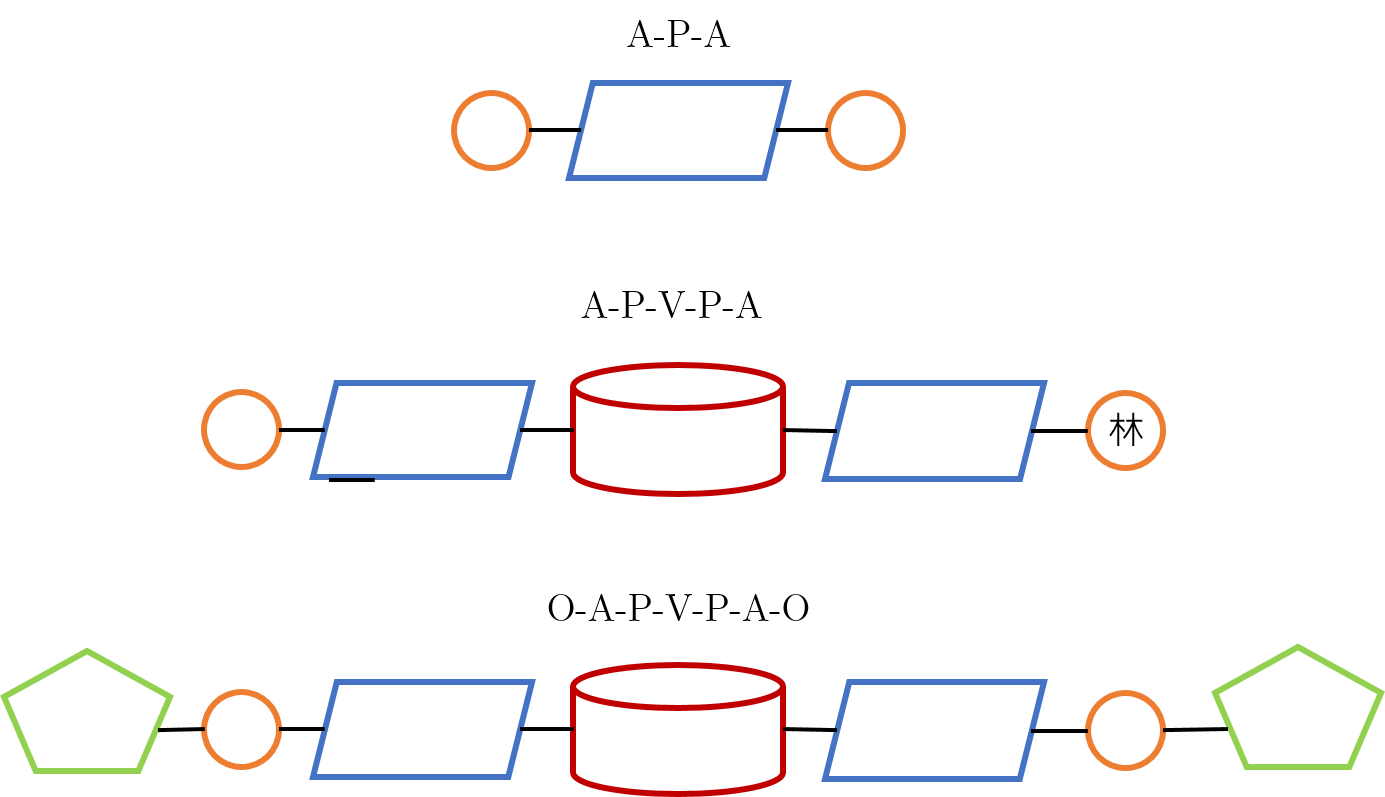}
    \caption{Metapaths for the academic network}
    \label{fig:metapaths}
\end{figure}
This work again uses the \skipgram{} algorithm, with the necessary adjustments for heterogeneous network. We are now looking to optimize 
\begin{equation}
    \argmax_\theta 
     \sum_{v \in V} \sum_{t \in T_V} \sum_{c_t \in N_t(v)} \log p(c_t|v;\theta)
\end{equation}
where $N_t(v)$ are the neighbors of $v$ of  the $t$th type. $p(c_t|v;\theta)$ is defined as the softmax function. 

As mentioned, the random walks must follow the meta-paths that are hand-designed for the specific network and task. Examples of meta-paths for the academic network are given in Fig. \ref{fig:metapaths}. For example, A-P-A denotes the Author-Paper-Author path. Thus, each specific meta-path semantic creates a bias toward specific relations. Some  meta-paths can be long, for example O-A-P-V-P-A-O (Organization-Author-Paper-Venue-Paper-Author-Organization). These meta-paths are then used to create biased random walks. I.e., the random walks must follow the semantics dictated by the various prescribed meta-paths. For example for A-P-A, the path with start with an author, then choose a paper (at random), then another author (at random). Again, this contexts are fed into a \skipgram{}-like neural network for the final embedding.

\subsection{Experiments and Results}

\begin{figure}
    \centering
    \includegraphics[width=3.5in]{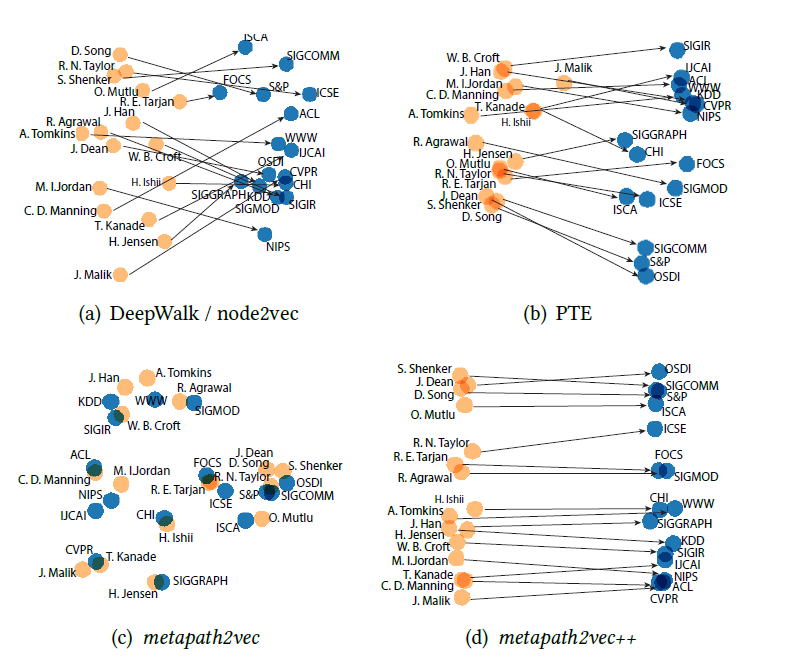}
    \caption{PCA projections of the 128D embeddings of 16 top CS conferences and corresponding high-profile authors \citep{dong2017metapath2vec}}
    \label{fig:mp-viz}
\end{figure}

\begin{figure}
    \centering
    \includegraphics[width=3.5in]{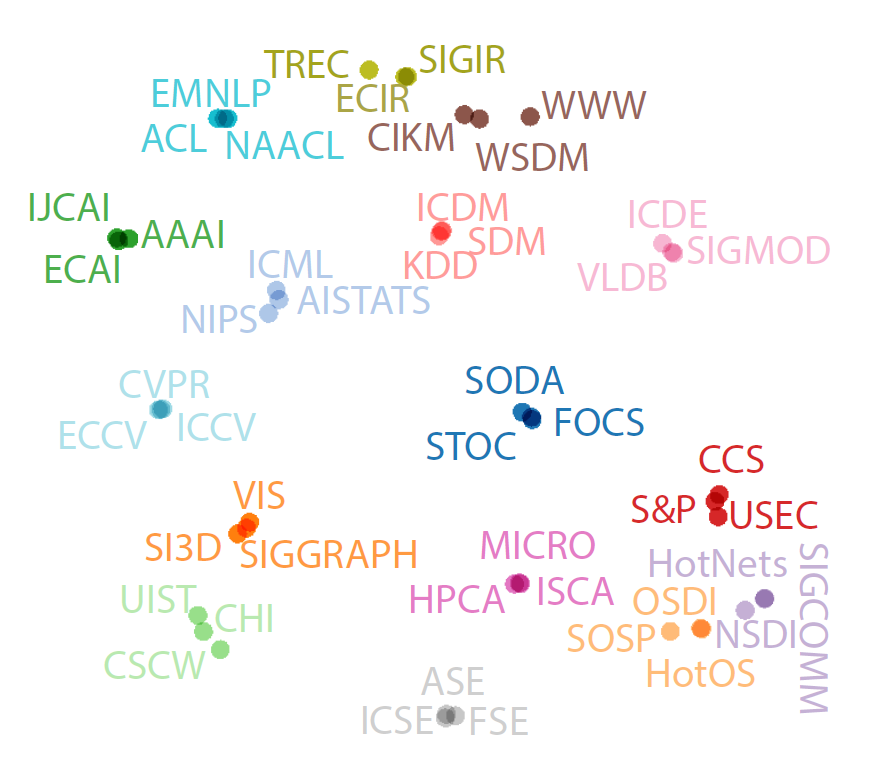}
    \caption{2D t-SNE projections of the 128D embeddings of
48 CS venues, three each from 16 sub-fields. \citep{dong2017metapath2vec}}
    \label{fig:mp-cluster}
\end{figure}

 The authors compared the \metapath{} performance to \deepwalk{}, LINE, and \nodevec{}, as well as to PTE (Predictive Text Embedding). The data used for the task consisted of two academic networks, AMiner Computer Science and  DBIS (Database and Information Systems). AMiner contains more than 9 million authors and 3 million papers from thousands of venues. DBIS is much smaller.
 
\metapath{} shows impressive results on tasks involving heterogeneous networks, including visualization, classification, and clustering. In Fig. \ref{fig:mp-viz}, PCA is used to project the embeddings of venues and top authors done by the various methods. The embeddings by \deepwalk{} (a) and PTE (b) clustered the authors and venues, but failed to create a meaningful relation between them. \metapath{} and \metapathplus{} (a variant of \metapath{}) shows a consistent relationship between each author and its field. 

Fig. \ref{fig:mp-cluster} demonstrates how the relations between nodes of a certain type also benefits from \metapath{}, as the projection of venues naturally lends itself to conferences in the same field being embedded close to each other. Performance in a multi-class classification task also showed superior performance compare to the baseline methods.

\section{Summary}
Methods of network embedding, which are based on research in word embeddings, are described.\footnote{Curiously, the \citet{perozzi2014deepwalk} paper never mentions the term \textit{network embedding} or \textit{graph embedding}, instead focusing on \textit{online learning} and \textit{representation}. The \citet{grover2016node2vec} paper mentions \textit{node embeddings} only once. This perhaps hints at disconnect between different research communities.}
By finding an analogy between documents and graphs, machine learning methods from NLP are successfully generalized to graphs. 
The methods, \deepwalk{}, LINE, \nodevec{}, etc. are able to process very large-scale graphs. Online learning of the latent vector representations of nodes are shown to have superior performance in various tasks, including multiple-label classification and edge prediction. Network embedding is an appropriate target for machine learning, since there are a multitude of underlying patterns in the graphs that are non-trivial to detect programmatically.

A major strength of these network embeddings is the ability to use well-developed data mining and other statistical algorithm (e.g., classification, prediction) for performing network tasks, instead of running discrete, path-, node-, or edge-based graph algorithms. The computational optimization are also an important advantage, as these methods scale to millions of nodes. Some of the issues which still need to be improved upon are the selection of hyper-parameters, and, perhaps most importantly, the interpretability of the embedded vectors. Similarly to word embeddings, while intuitively the method makes sense, there is currently no well-established theory that explains its success. 

Extending this research can be done in the direction using network embeddings for different heterogeneous networks, as well as experimenting with the embeddings in other network tasks, e.g. community detection, clustering, and bipartite graph analysis. Also, some of the insights gained from network embeddings (e.g., biased random walks, edge features) can potentially improve methods related to word embeddings back in the NLP domain. Another area for further research is creating  agreed-upon standard benchmarks for comparisons. Currently each paper develops its own experimental configuration. A development of standard benchmarks for network embedding thus has the potential to improve research and results in this area. This lack of standard benchmarks is probably a byproduct of the fact that there is no established theoretical framework for describing and analyzing the various methods and tasks. Finally, these methods are used in static networks. It would be interesting to try and expand methods to dynamic networks, in a way that would not require recalculating embeddings whenever a node or edge are added or removed from the network.

The field of machine-learning-based network embeddings continues to develop, with more techniques appearing regularly. 
Finally, the research evolution described, \wordvec{} (2013) $\rightarrow$ \deepwalk{} (2014) $\rightarrow$ LINE (2015) $\rightarrow$ \nodevec{} (2016), also demonstrates how quickly ideas propagate within the machine learning community, perhaps a testament to the success of the short-cycle conferences publication ecosystem. I also note that the code, data, and datasets for both the \deepwalk{} and \nodevec{} papers are available online. Thus, the methods and results can be tested, compared, reproduced, and improved upon. I see this as an essential asset for advancing research in this field.

\bibliographystyle{apalike}
\newpage
\bibliography{refs}

\end{document}